(76) Inventors: **Mark Davidson,** 807 Rorke Way, Palo Alto, CA (US) 94303
**Mario Rabinowitz,** 715 Lakemead Way, Redwood City, CA (US) 94062-3922

(57)            **ABSTRACT**

This invention deals with the broad general concept for focussing light. A mini-optics tracking and focussing system is presented for solar power conversion that ranges from an individual's portable system to solar conversion of electrical power that can be used in large scale power plants for environmentally clean energy. It can be rolled up, transported, and attached to existing man-made, or natural structures. It allows the solar energy conversion system to be low in capital cost and inexpensive to install as it can be attached to existing structures since it does not require the construction of a superstructure of its own. This novel system is uniquely distinct and different from other solar tracking and focussing processes allowing it to be more economical and practical. Furthermore, in its capacity as a power producer, it can be utilized with far greater safety, simplicity, economy, and efficiency in the conversion of solar energy.

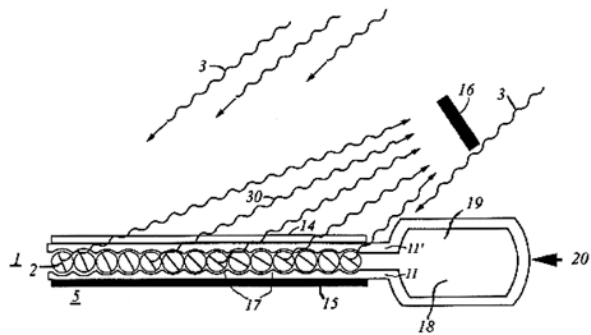

**28 Claims, 5 Drawing Sheets**

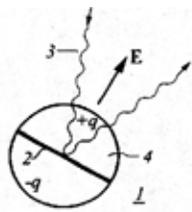
Fig. 1

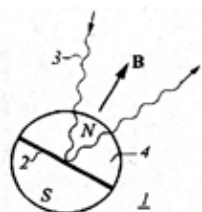
Fig. 2

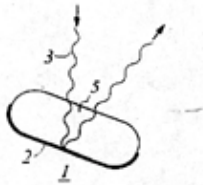
Fig. 3

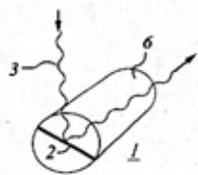
Fig. 4

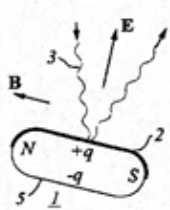
Fig. 5

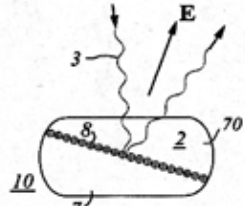
Fig. 6

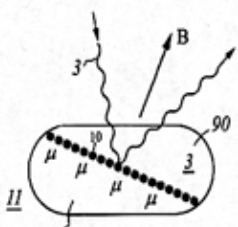
Fig. 7

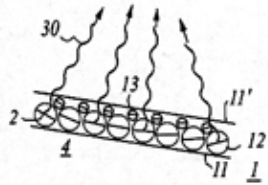
Fig. 8

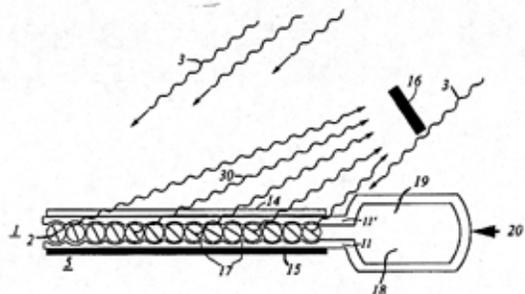
Fig. 9

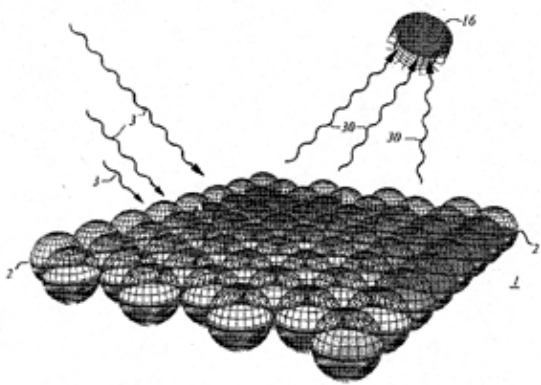

Fig. 10

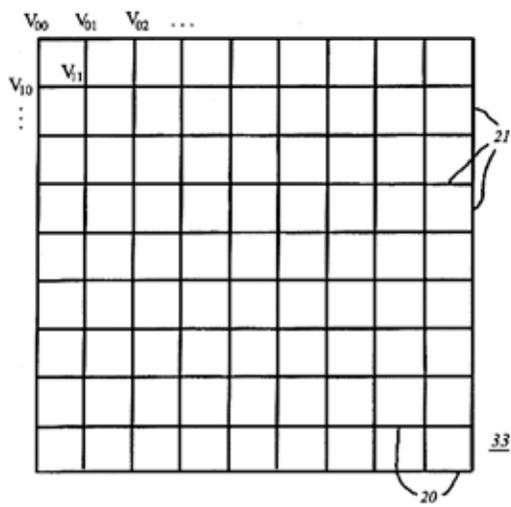

Fig. 11

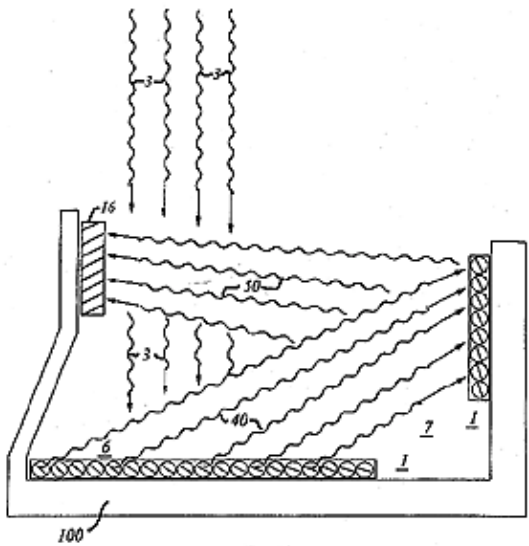

Fig. 12

# MINI-OPTICS SOLAR ENERGY CONCENTRATOR

BACKGROUND OF THE INVENTION
1. Field of the Invention

Due to an ever growing shortage of conventional energy sources, there is an increasingly intense interest in harnessing solar energy. A limiting factor in the utilization of solar energy is the high cost of energy converters such as photovoltaic cells. Our invention provides a low cost means for achieving affordable solar energy by greatly reducing the cost of solar concentrators which increase (concentrate) the density of solar energy incident on the solar energy con verter. For example, for the purpose of generating electricity, a large area of expensive solar cells may be replaced by a small area of high-grade photovoltaic solar cells operating in conjunction with the inexpensive intelligent mini-optics of our invention. Thus our invention can contribute to the goal of achieving environmentally clean energy on a large enough scale to be competitive with conventional energy sources.

Our invention is less expensive than conventional solar concentrators for two reasons. First due to miniaturization, the amount of material needed for the optical system is much *2S* less. Second, because our mini-optical solar concentrator is light-weight and flexible, it can easily be attached to existing structures. This is a great economic advantage over all existing solar concentrators which require the construction of a separate structure to support and orient them to intercept and properly reflect sunlight. Such separate structures must be able to survive gusts, windstorms, earthquakes, etc. The instant invention utilizes existing structureswhich are already capable of withstanding such inclement vicissitudes of nature.

2. Description of the Prior Art

There are many prior art patents that deal with twisting balls (gyricon) displays or separable balls displays. Electric or magnetic fields are used to orient or move these polarized or charged balls. To our knowledge none of the prior art utilizes the balls to optically concentrate (focus) light as in our invention. Furthermore the prior art neither teaches nor anticipates our application of the conversion of solar energy to electricity or any other form of energy. In one embodiment our invention incorporates balls with a shiny planar reflecting surface such as a metallic coating to give a high coefficient of reflectance. When the prior art refers to superior reflectance characteristics, they mean this in the context of displays with bi-colored balls e.g. black and white; or separable colored balls. In fact, the gyricon and separable so ball prior art do not teach the focussing of light in any capacity. These verities are evident from an examination of the prior art. A large representative sample of the prior art will now be enumerated and described. This together with the references contained therein constitutes a comprehensive compendium of the prior art.

U.S. Pat. No. 5,754,332 issued to J. M. Crowley on May 19, 1998 deals with gyricon bi-colored balls whose reflectance is comparable with white paper. Ile object is to produce a monolayer gyricon display.

U.S. Pat. No. 5,808,783 issued to J. M. Crowley on Sep. 15, 1998 deals with gyricon bi-colored balls "having superior reflectance characteristics comparing favorably with those of white paper." Again the objective is a display application.

U.S. Pat. No. 5,914,805 issued to J. M. Crowley on Jun. 22,1999 utilizes two sets of gyricon bi-colored balls "having superior reflectance charactreristics comparing favorably with those of white paper" for display purposes.

U.S. Pat. No. 6,055,091 issued to N. K. Sheridon and J. M. Crowley on Apr. 25, 2000 utilizes gyricon bi-colored cylinders. Again the objective is a display application.

U.S. Pat. No. 6,072,621 issued to E. Kishi, T. Yagi aud T. Ikeda on Jun. 6, 2000 utilizes sets of different mono-colored polarized balls which are separable for a display device.

U.S. Pat. No. 6,097,531 issued to N. K. Sheridon on Aug.1, 2000 teaches a method for making magnetized elements (balls or cylinders) for a gyricon display.

U.S. Pat. No. 6,120.588 issued to J. M. Jacobson on Sep. 19, 2000 describes a display device which uses mono-colored elements that are electronically addressable to change the pattern of the display.

U.S. Pat. No. 6,174,153 issued to N. K. Sheridon on Jan. 16, 2001 teaches apparatus for this purpose for a gyricon display.

U.S. Pat. No. 6,192.890 B1 issued to D. H. Levy and J.-P. F. Cherry on Feb. 27, 2001 is for a changeable tattoo display using magnetic or electric fields to manipulate particles in the display.

U.S. Pat. No. 6,211,998 BI issued to N. K. Sheridon on Apr. 3, 2001 teaches a method of addressing a display by a combination of magnetic and electric means.

U.S. Pat. No. 6,262,707 B I issued to N. K. Sheridon on Jul. 17, 2001 has a similar teaching for a gyricon display.

A large number of prior art devices have been described, all of which are directed at addressing and changing the pattern of a display device. While there are other such prior art teachings, none of them teaches or anticipates our invention.

## DEFINITIONS

"Bipolar" refers herein to either a magnetic assemblage with the two poles north and south, or an electric system with + and - charges separated as in an electret.

"Collector" as used herein denotes any device for the conversion of solar energy into other forms such as electricity, heat, pressure, concentrated light, etc.

"Compaction" refers to increasing the density of a collection (ensemble) of objects by geometrical arrangement or other means.

"Elastomer" is a material such as synthetic rubber or plastic, which at ordinary temperatures can be stretched substantially under low stress, and upon immediate release of the stress, will return with force to approximately its original length.

"Electret" refers to a solid dielectric possessing persistent electric polarization, by virtue of a long time constant for decay of charge separation.

"Electrophoresis or Electrophoretie, is an electrochemical process in which colloidal particles or macromolecules with a net electric charge migrate in a solution under the influence of an electric current. It is also known as catapboresis.

"Focussing planar mirroe, is a thin almost planar mirror constructed with stepped varying angles so as to have the optical *properties of* a much thicker concave (or convex) mirror. It can heuristically be thought of somewhat as the projection of thin equi-angular *segments of* small portions of a thick mirror upon a planar surface. It is a focusing planar reflecting surface much like a planar Fresnel lens is a focusing transmitting surface. The tracking-focussing property of an ensemble of tiny elements which make up the focussing planar mirror are an essential feature of the instant invention.

"Hehostat" denotes a clock-driven mounting for automatically and continuously pointing apparatus in the direction of the sun.

"Immiscible" herein refers to two fluids which are incapable of mixing.

"Packing fraction" herein refers to the fraction of an available volume or area occupied by a collection (ensemble) of objects.

"Polar gradient" as used herein relates to magnetic optical elements that are controlled in the non-gyricon mode such as in the magnetic field gradient mode.

"Monopolar" as used herein denotes mono-charged optical elements that are controlled in the non-gyricon mode such as the electrophoretic mode.

"Rayleigh limit" relates to the optical limit of resolution which can be used to determine the smallest size of the elements that constitute a mini-mirror. Lord Rayleigh discovered this limit from a study of the appearance of the diffraction patterns of closely spaced point sources.

"Spin glass" refers to a wide variety of materials which contain interacting atomic magnetic moments. They possess a form of disorder, in which the magnetic susceptibility undergoes an abrupt change at what is called the freezing temperature for the spin system.

"Thermoplastic" refers to materials with a molecular structure that will soften when heated and harden when cooled. This includes materials such as vinyls, nylons, elastomers, fuorocarbons, polyethylenes, styrene, acrylics, cellulosics, etc.

"Translucent" as used herein refers to materials that pass light of only certain wavelengths so that the transmitted light is colored.

## SUMMARY OF THE INVENTION

There are many aspects and applications of this invention. Primarily this invention deals with the broad general concept of method and apparatus for focussing light. A particularly important application is the focussing of sunlight for power conversion and production.

It is a general object of this invention to provide a focussing planar mini-optic system for reflecting light with a substantially higher power density than the incident light.

One object is to provide an inexpensive, light-weight, and flexible mini-optical light concentrator that can easily be attached to existing structures, and thus does not require the construction of a superstructure of its own.

Another objective is to provide a solar energy conversion system that is not only low capital cost, but that is also inexpensive to install.

A particularly important object is to provide a unique tracking and focussing system for solar power conversion.

Another object is to provide a system that holds or locks the mini-mirror elements in rigid orientation with minimal to no-power expenditure between rotational focussing operations.

Another object is to provide a means for unlocking the mini-mirror elements so that they may rotate freely when being guided into the proper orientation.

Another object is to provide an inexpensive system for photovoltaic conversion.

Another objective is to provide daily peaking power when the load is highest on the conventional power grid.

Another objective is to provide electricity to remote villages or rural settlements.

Another object is to provide a rugged system for conversion of solar energy to heat.

Another objective is to provide electricity for communications installations.

Another object is to provide large-scale environmentally clean energy.

Another objective is to help in the industrialization of developing countries.

Another object is to provide a low-cost, tough, lightweight, concentrated efficient solare energy converter that is highly portable.

Another objective is to provide a minitiarized planar heliostat field configuration that can either track the sun temporally, or follow the sun with a photomultiplier which searches for a maximum output.

Another object is to provide a portable system that can easily go anywhere man can go, to track and concentrate the sun's energy.

Other objects and advantages of the invention will be apparent in a description of specific embodiments thereof, given by way of example only, to enable one skilled in the art to readily practice the invention as described hereinafter with reference to the accompanying drawings.

In accordance with the illustrated preferred embodiments, method and apparatus are presented that are capable of producing and maintaining a high concentration of light relative to the original source such as sunlight. The embodiments are all capable of secure attachment to sturdy existing structures to provide an inexpensive application with a long and trouble-free life.

### BRIEF DESCRIPTION OF THE DRAWINGS

FIG. 1 is a cross-sectional view of an electrically charged bipolar sphere with an equatorial flat reflecting surface. This sphere is one of a multitude of optical elements which track the sun and focus the sun's light beam onto a collector.

FIG. 2 is a cross-sectional view of a magnetically charged bipolar sphere with an equatorial flat reflecting surface. This sphere is one of a multitude of optical elements, which track the sun and focus the sun's light beam onto a collector.

FIG. 3 is a cross-sectional view of a circular disk with a backside-reflecting surface. This disk is one of a multitude of optical elements which track the sun and focus the sun's light beam onto a collector.

FIG. 4 is a cross-sectional view of a cylinder with an internal flat reflecting surface. This cylinder is one of a multitude of optical elements which track the sun and focus the sun's light beam onto a collector.

FIG. 5 is a cross-sectional view of a circular disk with a frontside reflecting surface. This disk is one of a multitude of optical elements which track the sun and focus the sun's light beam onto a collector.

FIG. 6 is a cross-sectional view of a monopolar electric cell filled with two immiscible fluids, and shiny charged particles of the same sign in the bottom one. 'Ibis cell is one of a multitude of optical elements which track the sun and focus the sun's light beam onto a collector.

FIG. 7 is a cross-sectional view of a ferrofluid cell partially filled with a colloidal suspension of shiny ferromagnetic particles in a fluid. This cell is one of a multitude of optical elements which track the sun and focus the sun's light beam onto a collector.

FIG. 8 is a cross-sectional view of a mini-optics ensemble of elements of two or more populations of sizes to increase the packing fraction and hence the reflectance. Each element tracks the sun and focuses the sun's light beam onto a collector.

FIG. 9 is a cross-sectional view of a mini-optics ensemble of elements showing the overlay of a transparent ground plane on top and a resistive grid on the bottom to locally produce varying mini-electric fields for orienting the mini-mirrors to focus the incident light onto a collector.

FIG. 10 is a perspective view of a two-dimensional array of the rotatable elements of a focussing planar mirror.

FIG. 11 is a schematic top view showing the electronic control grid for rotating the reflecting elements of a focussing planar mirror.

FIG. 12 illustrates method and apparatus for significantly increasing the degree of concentration of solar energy reaching the collector by utilizing two or more focussing planar mirrors.

### DETAILED DESCRIPTION OF THE PRESENTLY PREFERRED EMBODIMENTS

FIG. 1 shows a rotatable element 1 of a focussing planar mini-mirror with an equatorial flat reflecting surface 2 which 25 reflects a wave beam of sunlight 3. The element I shown is a cross-sectional view of an electrically charged bipolar sphere 4 with charge +q in one hemisphere and charge -q in the opposite hemisphere, that is operated in the well-known electrical gyricon mode. This sphere 4 is one of a multitude of rotatable optical elements 1 which track the sun and focus the sun's light wave beam onto a collector by means of an electric field E.

FIG. 2 shows a rotatable element 1 of a focussing planar mini-mirror with a flat equatorial reflecting surface 2 which reflects a wave beam of sunlight 3. The element 1 shown is a cross-sectional view of a magnetically charged bipolar sphere 4 with north magnetic field N in one hemisphere and south magnetic field S in the other hemisphere, that is operated in the well-known magnetic gyricon mode. This sphere 4 is one of a multitude of rotatable optical elements 1 which track the sun and focus the sun's light beam onto a collector by means of a magnetic field B.

FIG. 3 shows a rotatable element 1 of a focussing planar mini-mirror with a backside reflecting surface 2 which reflects a light wave beam 3. The element I shown is a cross-sectional view of a circular disk 5 with rounded edges, that is operated in any of the well-known modes, such as gyricon, electrical monopolar, magnetic, polar gradient, etc. Ibis disk 5 is one of a multitude of rotatable optical elements 50 1 which track the sun and focus the sun's light wave beam onto a collector by means of an electric field or magnetic field, or combination thereof. It should be noted that in display modes, a spherical or cylindrical shape is necessary for the elements, as they must be able to rotate 180 degrees without binding up in order to display a black or white side up. In the instant invention, a 90 degree rotation of the element I is more than sufficient as this produces a 180 degree reflection of the beam of sunlight. Since the angle of reflection is equal to the angle of incidence on the reflecting element 1, a doubling of the angle is produced.

FIG. 4 shows a rotatable element I of a focussing planar mini-mirror with an internal flat reflecting surface 2 in the plane of the hemicylinders which reflects a wave beam of sunlight 3. The element 1 shown is a cross-sectional view of a cylinder 6 that is operated in any of the well-known modes, such as gyricon, electrical monopolar, magnetic, polar gradient, etc. This cylinder 6 is one of a multitude of rotatable optical elements I which track the sun and focus the sun's light beam onto a collector by means of an electricfield or magnetic field, or combination thereof.

In the case of non-front-surface reflection such as shown in FIGS. 1-4, the material of element 1 needs to be clear or transparent so the incident light can easily reach the reflecting surface 2.

FIG. 5 shows a rotatable element 1 of a focussing planar mini-mirror with a frontside reflecting surface 2 which reflects a light wave beam 3. This is a presently preferred embodiment of the rotatable element 1. The element I shown is a cross sectional view of a circular disk 5, with rounded edges that is operated in any of the well-known modes, such as gyricon, electrical monopolar, magnetic, polar gradient, etc. Ibis disk 5 is one of a multitude of optical elements 1 which track the sun and focus the sun's light wave beam onto a collector by means of an electric field E or magnetic field B, or combination thereof. The case is illustrated where two-axis control is possible in mutually orthogonal directions by means of embedded charge +q and -q at top and bottom, and embedded magnetic field with north magnetic field N at one end and south magnetic field S in the other as shown. TWo-axis control can also be accomplished with either an E or B field singly.

It should be noted that in prior art display modes, a spherical or cylindrical shape is necessary for the elements, as they must be able to rotate 180 degrees without binding up in order to display a black or white side up. In the instant invention, a 90 degree rotation of the element I is more than sufficient as this produces a 180 degree reflection of the beam of sunlight, since the angle of reflection is equal to the angle of incidence on the reflecting element 1. *Thus* a doubling of the angle is produced herein.

FIG. 6 shows a fixed element 10 of a focussing planar mini-miffor which is a cross-sectional view of a monopolar electric cell 2 partially filled with a bottom fluid 7 with shiny charged particles 8 of the same sign (shown here as +, but which could also all be -), and a top transparent fluid 70. The two fluids are immiscible. When an electromagnetic field E' is applied, the particles 8 coalesce to form a flat reflecting surface at the interface between fluid 7 and fluid 70, as also influenced by surface tension and meniscus. Fluid 70 could be air, but a transparent fluid of substantially less density than fluid 7 is preferred so that gravity will act to maintain their relative top/bottom orientations. If the particles 8 are small enough to form a colloidal suspension, the density of the particles 8 and the fluid 7 may differ. However, it is generally preferable to have the density of the particles 8 approximately matched to the fluid 7.

The orientation of this flat reflecting surface can be controlled by E to reflect light 3. Until the electric field E is applied, as an optional capability the particles 8 and the fluid 7 can function as a transparent window when the particles 8 are nanosize i.e. much smaller than the wavelength of the incident light and the fluid 7 is transparent or translucent while they are dispersed in the fluid 7. For the case of dispersed transparency' the particles 8 should be <<4000 A *(4x 10-7 in). This* cell 2 is one of a multitude of optical elements I which track the sun and focus the sun's wave beam onto a collector. 7be particles 8 may include a wide variety of electomagnetically interactive materials such as electret, optoelectric, conducting, thermoelectric, electrophoretic, resistive, serniconductive, insulating, piezoelectric, magnetic, ferromagnetic, paramagnetic, diamagnetic, or spin (e.g. spin glass) materials. It should be noted that the reflecting area remains constant for spherical and circular-cylindrical cells, as the orientation of the reflecting surface changes. However, the increase in reflecting area is not a serious problem for the non-spherical, non-4ircular cell geometry shown.

FIG. 7 shows a fixed element 11 of a focussing planar mini-mirror which is a cross-sectional view of a ferrofluid cell 3 partially filled with a ferrofluid 9 containing shiny ferromagnetic particles 10 of high permeability, shown here as u, and a top transparent fluid 90. The two fluids are immiscible. When an inhomogeneous electromagnetic field B of increasing gradient is applied, the particles 10 are drawn to the region of increasing gradient and coalesce to form a flat reflecting surface at the interface between fluid 9 and fluid 90, as also influenced by surface tension and meniscus. Fluid 90 could be air or a transparent fluid of substantially less density than fluid 9 so that gravity will act to maintain their relative top/bottom orientations. The orientation of the flat reflecting surface can be controlled by B to reflect light 3. This cell 3 is one of a multitude of optical elements I which track the sun and focus the sun's wave beam onto a collector. The particles 10 are small enough to form a colloidal suspension, and are coated to prevent coalescence until B is applied, as is well known in the art. It should be noted that the reflecting area remains constant 20 for spherical and circular-cylindrical cells, as the orientation of the reflecting surface changes. However, the increase in reflecting area is not a serious problem for the non-spherical, non-circular cell geometry shown.

FIG. 8 is a cross-sectional view of a mini-optics ensemble 4 of rotatable elements 1 of two or more populations of particle sizes to increase the packing fraction and hence increase the energy of the reflected wave 30. The particles are contained between two elastomer sheets 11 of which the top sheet 11' is transparent. The large particles 12 and the small particles 13 can already be rotatable, or rendered rotatable by expanding the clastomers. 11 by the application of a fluid thereto. TU small particles 13 are disposed in the interstices of the monolayer arrangement of the large particles 12. Thus the small particles 13 just fit into the small pockets created by the conjunction of the large particles 12, to create more reflecting area than the very small area that these small particles 13 block of the large particles 12. Each element I tracks the sun and focuses the sun's light beam onto a collector.

Let us here consider the packing (compaction) of spheres in broad terms so that we may better understand the various trade-offs that may be undertaken in the choice of one set of particles 12 versus two sets of particles 12 and 13, or more; and the relative advantages that are also a function of the packing array. (The spheres are chosen for convenience. We could equally well be discussing circular disks as in FIGS. 3 and 5) With one set of particles 12 of radius R in a square monolayer array in which any adjacent four particles have their centers at the corners of a square, the maximum packing fraction of the circular mirrors is

$$PFs1 = \frac{4(\pi R^2)}{(4R)^2} = \frac{\pi}{4} = 0.785.$$

This means that as much as 21% the reflecting area is wasted, with less than 79% of the area available for reflection. If a second population of particles 13 are put into the interstices, their radii would need to be just slightly greater than

$$r_s > R[\sqrt{2} - 1] = 0.414\ R$$

so that they would fill the interstices of a monolayer of spheres (first population), and yet not fall through the openings. The maximum packing fraction in square array of two such sets of circular mirrors is

$$PFs2 = \frac{4(\pi R^2) + \pi r^2 + 4\left(\frac{1}{2}\pi r^2\right) + 4\left(\frac{1}{4}\pi r^2\right)}{(4R)^2} = 0.920.$$

Thus just by the addition of a second population of particles 13, of the right size, the reflecting area can increase from about 79% to about 92% in a square array.

Now lot us consider one set of particles 12 of radius R in a hexagonal monolayer array in which any adjacent six particles have their centers at the corners of a hexagon. In this case, the maximum packing fraction of the circular is mirrors is

$$PFh1 = \frac{\pi R^2 + 6\left(\frac{1}{3}\pi R^2\right)}{6\left[R(R\sqrt{3})\right]} = 0.907.$$

This means that only about 10% the reflecting area is wasted, with about 90% of the area available for reflection with one population of particles 12, by just going to a hexagonal array. If a second population of particles 13 are put into the interstices, their radii would need to be just slightly greater than

$$r_h > R[\tfrac{2}{3}\sqrt{3}-1] = 0.155R$$

so that they would fill the interstices of a monolayer of an hexagonal affay of spheres (first population of particles 12), and yet not fall through the openings. The maximum packing fraction in hexagonal array of two such sets of circular mirrors is

$$PFh2 = \frac{\pi R^2 + 6\left(\frac{1}{3}\pi R^2\right) + 6\pi r^2}{6\left[R(R\sqrt{3})\right]} = 0.951.$$

Thus just by the addition of a second population of particles 13, of the right size, the reflecting area can increase from about 90% to about 95% in an hexagonal array.

The following two tables summarize the above results on packing fractions.

### TABLE 1

#### Comparison of Hexagonal and Square Packing Fractions

|  | PF1 | PF2 | PF2/PF1 |
|---|---|---|---|
| Hexagonal Packing | 0.907 | 0.951 | 1.049 |
| Square Packing | 0.785 | 0.920 | 1.172 |

### TABLE 2

#### Relative Gain of Hexagonal versus Square Packing

| PFh1/PFs1 | PFh2/PFs2 | PFh2/PFs1 |
|---|---|---|
| 1.155 | 1.034 | 1.211 |

Interesting conclusions can be drawn from TABLES 1 and 2 which can be guides for design tradeoffs even though the calculated quantities are upper limits of what can be attained in practice. TABLE 2 shows that just by going from a square monolayer array to an hexagonal monolayer array the reflecting area can be increased by about 16%. When two populations of particles 12 and 13 are used, there is only about a 3% improvement by going to an hexagonal array. The largest improvement is about 21% for a two population hexagonal array compared with a one population square array.

FIG. 9 is a cross-sectional view of a mini-optics ensemble 5 of an individually rotatable monolayer of elements I showing the overlay of a transparent ground plane 14 on top and a resistive grid 15 on the bottom to locally produce varying mini-electric fields for orienting the mini-mirrors 2 to focus the incident light 3 as concentrated light of the reflected wave 30 onto a collector 16. The collector 16 as used herein denotes any device for the conversion of solar energy such as electricity, heal, pressure, concentrated light, etc. The rotatable elements 1 are situated in ridged cells 17 *15* between two elastomer sheets. For spherical or cylindrical elements A the ridged cellular structure 17 is conducive but not necessary to hold the elements in grid position in the array structure. For elements 1 of disk shape 5 as in FIGS. 3 and 5, the ridged cells 17 are a valuable adjunct in maintaining the array structure and avoiding binding between the elements 1.

Because the mini-optics system is tough and light-weight it is highly portable unlike existing light concentrating optical systems that are heavy, bulky, and cumbersome. Furthermore, the mini-optics system can easily be mass produced at low cost since it is mainly two sheets of thin plastic with millions of smart-beads sandwiched between the sheets. The mini-optics system can be rolled up, transported, and attached to existing man-made, or natural structures such as trees, rocks, hillsides, and mountain tops. Therefore, in addition to providing solar energy in conventional urban settings, the mini-optics system is also ideally suited for rugged terrain and can be used by campers, mountain climbers, explorers, etc. It is unmatched as a portable system that can easily go anywhere man can go, and track and concentrate the sun's energy.

When rotation of the elements I is desired, the effect of the torque applied by the field can be augmented by injecting a fluid 18 from a plenum reservoir 19 by a pressure applying means 20 to expand the separation of the sheets 11. In the case of non-front-surface reflection such as shown in FIGS. 1-4, it is desirable to utilize a fluid 18 whose index of refraction matches the clear hemisphere or clear hemicylinder. In addition to providing a means to pressure the elastomer sheets 11 apart, the fluid 18 acts as a lubricant to permit the elements 1 to rotate freely when being guided into the proper orientation.

The ridged cells 17 can be created in thermoplastic elastomer sheets 11 by heating the sheets 11 to a slightly elevated temperature and applying pressure with the elements 1 between the sheets 11. In the case of elements 1 of disk shape 5, the ridged cells 17 can be created on each sheet individually. This gives twice the height for the cells, when two such sheets are put together to hold the elements 1.

A presently preferred maximum for the diameter of elements 1 is -10 mm or more for this figure and for FIGS. 1-5. The minimum diameter of elements I can be assessed from the Rayleigh limit

$$d = \frac{0.61\lambda}{n \sin u} \sim 10\lambda,$$

where d is the minimum diameter of elements 1, $\lambda \sim 4000$ A is the minimum visible wavelength, n is the index of refraction -1 of element I (the medium in which the incident light is reflected), and u is the half angle admitted by elements 1. Thus $d \sim 40,000$ A ($4 \times 10^{-6}$ m) is the minimum diameter of elements 1.

If the focussing planar mini-mirrors concentrate the solar radiation by a factor of 100, the total increase in power density reaching the collector would be 100 times greater than the incident power of the sun. Thus the collector area need be only ~ 1% the size of one receiving solar radiation directly. Although the total capital and installation cost of this improved system may be more than 1% of a direct system, there will nevertheless be substantial savings.

FIG. 10 is a perspective view of a two-dimensional array of the rotatable elements 1 of a focussing planar mini-mirror with an equatorial flat reflecting surface 2 which reflects incident light 3 and focuses it as a concentrated light wave 30 to a collector 16.

FIG. 11 is a schematic top view showing the electronic control grid 33 for rotating the reflecting elements of a focussing planar mini-mirror. Except for the cylinders of FIG. 4 which have a one-axis response, the preferred non-cylindrical geometry of each of the other elements 1, 10, or 11 is capable of rotating in any direction (two-axis response) in response to a selectively applied electric field by the electronic control grid 33. The electronic control grid 33 is made of resistive components 21. The mini-mirror/lens array with elements 1, 10, or 11 is sandwiched between the resistive electronic control grid shown here and the transparent ground plane as shown in the cross-sectional view of FIG. 9. The orientation of the elements 1, 10, or 11 (cf. FIGS. 1-7) is determined by controlling the voltages V at the nodes of the grid such as those shown $V_{00}$, $V_{01}$, $V_{02}$, $V_{10}$, $V_{11}$, with voltage $V_{ij}$ at the ij th node. The voltage $V_{ij}$ can be controlled by a very small inexpensive computer with analog voltage outputs. Once in operation, this system can be powered by the solar energy conversion device which collects the concentrated light. The electronic control grid 33 is similar in construction and function to analogous grids used in personal computer boards, and in flat panel monitors.

The voltage between successive nodes produces an electric field in the plane of the planar mini-mirror, and the voltage between a node and the ground plane produces an electric field perpendicular to the planar mini-mirror to control the orientation angle of the reflecting/focussing mini-mirrors. The number of elements 1, 10, or 11 per grid cell is determined by the degree of focussing desired: the higher the degree of focussing, the fewer the number of elements per grid cell. In the case of elements 1 which contain orthogonal electrical and magnetic dipoles as in FIG. 7, the orientation function may be separated for orientation in the plane and orientation perpendicular to the plane by each of the fields.

After being positioned for optimal focussing angles of reflection, elements 1 (cf. FIGS. 1-5, and 9) may be held in place by the elastomer sheets 11 (cf. FIGS. 8 and 9) with the voltages $V_{ij}$ being turned off to eliminate unnecessary power dissipation. When a new angular orientation of the elements I and 2 is desirable due to the sun's motion relative to the earth, the sheets 11 (cf. FIG. 9) are separated by injecting a fluid 18 from a plenum reservoir 19 by a pressure applying means 20. In the case of elements 10 or 11 the reflecting angle needs to be held fixed by the control function which is the electronic control grid 33. To minimize power dissipation in this case it is desirable to make resistive components 21 highly resistive so that a given voltage drop is accomplished with a minimum of current flow and hence with a minimum of power dissipation.

FIG. 12 illustrates method and apparatus for significantly increasing the degree of concentration of solar energy reaching the collector by utilizing two or more focussing planar mini-mirrors. Shown are a cross-sectional view of two sets of mini-optics ensembles 6 and 7 of rotatable elements I wherein the sunlight 3 is incident on the first ensemble 6 and the reflected light 40 from this first ensemble 6 is focussed on the second ensemble 7 to reflect light 50 which is further concentrated and focussed on the collector 16. The mini-optics ensembles 6 and 7 are attached to a structure 100 which preferably is a pre-existing structure such as a building.

To illustrate the magnification capability of this configuration, in the ideal case where all the incident light is reflected without absorption or losses, if the two sets of focussing planar mini-mirrors each concentrated the light energy by a factor of 100, the total increase in power density reaching the collector would be a factor of (100)' -10,000 times greater than the incident power. For n such reflectors each feeding into the other until finally reaching the collector, the increase would be (100r. Similarly, if two focussing planar mini-mirrors were positioned to have n concentrating reflections between them before the light is reflected to the collector, the increase would also be (100r. Of course in a real case the increase would be less than this due to losses. The thermodynamic limit of such a scheme would be an effective temperature of the radiation at the collector no higher than the source temperature which in the case of the sun is ~ 6000 K. A practical limit would occur much before this related to temperatures well below the melting point of the materials used. There is also an optical limit that the power per unit area per steradian cannot be increased by a passive optical system.

A major advantage of the instant invention, is that the focussing planar mini-mirrors can be attached to an already existing structure 100, ranging from buildings to trees t; rocks, or even flat on the ground. Since the necessary structural strength is already built into the existing structure, no additional funds need be expended to construct mirror supporting edifices which are capable of withstanding harsh winds, earthquakes, and other natural disturbances.

While the instant invention has been described with reference to presently preferred and other embodiments, the descriptions are illustrative of the invention and are not to be construed as limiting the invention. Thus, various modifications and applications may occur to those skilled in the art without departing from the true spirit and scope of the invention as summarized by the appended claims.

What is claimed is:

1. A miniature optics system for concentrating reflected sunlight, comprising:
   (a) an array of miniature rotatable reflecting balls having two hemispheres positioned in the space between two sheets holding said array of miniature reflecting balls;
   (b) the top sheet of said two sheets being transparent;
   (c) means to individually rotate the array of balls within said sheets; and
   (d) a dipole embedded in each ball for coupling to at least one of the sheets.
2. The apparatus of claim 1, wherein each said rotatable miniature reflector is a ball comprising:
   (a) a reflector embedded in said ball; and
   (b) charge of opposite sign in each said ball.
3. The apparatus of claim 1, wherein each said rotatable miniature reflector is a ball comprising:
   (a) a reflector embedded in said ball; and
   (b) a magnetic dipole each said ball.
4. The apparatus of claim 1, wherein each said rotatable miniature reflector is a disk comprising:

(a) a reflector on one of the surfaces of said disk; and
   (b) bipolar charge in each disk.
5. The apparatus of claim 1, wherein each said rotatable miniature reflector is a disk comprising:
   (a) a reflector on one of the surfaces of said disk; and
   (b) a magnetic dipole embedded in said disk.
6. The apparatus of claim 1, wherein each said rotatable
   miniature reflector is a disk comprising:
   (a) a reflector on one of the surfaces of said disk, and
   (b) a magnetic dipole and an electric dipole embedded in said disk.
7. The apparatus of claim 1, wherein each said rotatable
   miniature reflector is a cylinder comprising:
   (a) a reflector embedded in said cylinder; and
   (b) means for coupling to an electomagnetic field.
8. The apparatus of claim 1, wherein said sheets maintain
   said reflectors in fixed azimuthal orientation.
9. The apparatus of claim I with means for spreading
   apart said sheets.
10. The apparatus of claim 1, wherein the diameter of each said rotatable miniature reflector is in
    the range $4 \times 10^{-6}$ m to $10^{-1}$ m .
11. The apparatus of claim 1, wherein at least one sheet forms a cellular array.
12. The apparatus of claim 1, wherein said concentrating
    reflected light is caused to produce power.
13. The apparatus of claim 1, wherein said concentrating reflected light is caused to produce
    electrical power.
14. A miniature optics system for concentrating reflected sunlight, comprising:
    (a) an array of rotatable miniature reflectors positioned in
       the space between two sheets;
    (b) means for rotating said rotatable miniature reflectors;
    (c) means for tracking the source of light;
    (d) means for focusing said reflecting system unto a collector; and
    (e) a dipole embedded in each reflector for coupling to at
       least one of the sheets.
15. The apparatus of claim 14, wherein each said rotatable
    miniature reflector is a body comprising:
    (a) a reflector embedded in said body; and
    (b) field producing sources in each quadrant of said body.
16. The apparatus of claim 14, wherein the said sheets maintain the reflectors in fixed azimuthal
    orientation.
17. The apparatus of claim 14 with means for spreading
    apart the sheets.
18. The apparatus of claim 14, wherein said concentrating reflected light is caused to produce
    power.
19. The apparatus of claim 14, wherein said concentrating reflected light is caused to produce
    electrical power.
20. The apparatus of claim 14 wherein the concentrated light is used in a process to charge a fuel
    cell.
21. The apparatus of claim 14 wherein the concentrated light is used in a process to de-salinate
    water.

22. The apparatus of claim 14 wherein the concentrated light is used in a process to charge batteries.
23. The apparatus of claim 14 wherein the concentrated light is used in a process to produce hydrogen fuel.
24. The apparatus of claim 14 wherein the concentrated light is used to heat a building.
25. The apparatus of claim 14 wherein the concentrated light is used to melt snow or ice.
26. The apparatus of claim 14 wherein the concentrated light is used to heat water.
27. The apparatus of claim 14 wherein the concentrated light is used to provide heat for a distillation process.
28. The apparatus of claim 14 wherein the concentrated light is used to provide heat or electricity for powering an ocean-going vessel.